\documentclass[preprint,review,12pt]{elsarticle}

\usepackage{amssymb,amsmath,lineno}

\journal{arXiv}

\begin{document}

\begin{frontmatter}

\title{Exact sharp-fronted solutions for nonlinear diffusion on evolving domains}

\author{Stuart T. Johnston\corref{correspondingauthor}}
\address{School of Mathematics and Statistics, The University of Melbourne, Victoria, Australia.}
\cortext[correspondingauthor]{Corresponding author}
\ead{stuart.johnston@unimelb.edu.au}

\author{Matthew J. Simpson}
\address{School of Mathematical Sciences, Queensland University of Technology, Brisbane, Australia.}

\begin{abstract}
Models of diffusive processes that occur on evolving domains are frequently employed to describe biological and physical phenomena, such as diffusion within expanding tissues or substrates. Previous investigations into these models either report numerical solutions or require an assumption of linear diffusion to determine exact solutions. Unfortunately, numerical solutions do not reveal the relationship between the model parameters and the solution features. Additionally, experimental observations typically report the presence of sharp fronts, which are not captured by linear diffusion. Here we address both limitations by presenting exact sharp-fronted solutions to a model of degenerate nonlinear diffusion on a growing domain. We obtain the solution by identifying a series of transformations that converts the model of a nonlinear diffusive process on an evolving domain to a nonlinear diffusion equation on a fixed domain, which admits known exact solutions for certain choices of diffusivity functions. We determine expressions for critical time scales and domain growth rates such that the diffusive population never reaches the domain boundaries and hence the solution remains valid.
\end{abstract}

%


\begin{keyword}
Diffusion \sep Exact solution \sep Growing domain \sep Sharp fronts
\end{keyword}

\end{frontmatter}


\section{Introduction}
\noindent Spatial domains that undergo growth are ubiquitous in the natural world \cite{bers2002,landman2007,riess1998,shraiman2005}. Accordingly, many (bio)physical applications involve diffusive processes that occur on growing domains \cite{bers2002,landman2007}. There is significant interest in developing exact solutions to models of such processes \cite{abad2020,johnston2023,le2018,ryabov2015,simpson2015b,simpson2015c,yuste2016}. Previous models either rely on an assumption of linear diffusion \cite{simpson2015b,simpson2015c,yuste2016} or require numerical solutions \cite{abad2020,le2018}. Models with linear diffusion do not produce solution profiles with sharp fronts, which are observed frequently in experimental data \cite{garduno1994,mccue2019}, while numerical solutions do not provide explicit relationships between model parameters and solution behaviour. Here we simultaneously relax both restrictions. \\

A common approach employed to model diffusion on a growing domain considers a $d$-dimensional radially-symmetric growing spatial domain defined according to $-L(t) \leq r \leq L(t)$, where $L(t) > 0 \ \forall t$ \cite{simpson2015b}. The evolution of a population with density $C(r,t)$ that undergoes diffusive transport on this growing domain according to the nonlinear diffusivity function $D(C)$ is
\begin{align}
\frac{\partial C(r,t)}{\partial t} &= \frac{1}{r^{d-1}}\frac{\partial}{\partial r}\bigg( r^{d-1}D(C)\frac{\partial C(r,t)}{\partial r}\bigg) -  \frac{1}{r^{d-1}}\frac{\partial}{\partial r}\bigg(r^{d-1}v(r,t)C(r,t)\bigg), 
\label{Eq:GoverningMain} \\  & -L(t) < r < L(t), \ C(r,0) = F(r) \nonumber.
\end{align}
The domain undergoes growth due to the expansion of each infinitestimal width of space according to
\begin{equation*}
\frac{\text{d}L(t)}{\text{d}t} = \int_0^{L(t)} \frac{\partial v(r,t)}{\partial r} \ \text{d}r,
\end{equation*}
where $v(r,t)$ is the velocity field that translates each point in space. To facilitate analysis, it is standard to assume that the domain undergoes spatially uniform growth (i.e. $\partial v/\partial r$ is independent of position)
\begin{equation*}
\frac{\partial v(r,t)}{\partial r} = \sigma(t) = \frac{1}{L(t)}\frac{\text{d}L(t)}{\text{d}t},
\end{equation*}
and hence
\begin{equation*}
v(r,t) = \frac{r}{L(t)}\frac{\text{d}L(t)}{\text{d}t}.
\end{equation*}
The diffusive component of the governing equation can be considered to represent the random motion of the population, while the advective component can be considered to represent directed motion of the population due to the expansion of space. Previous exact solutions of this model focus on the linear diffusion case (i.e. $D(C) = D$); with investigations into the role of dimensionality \cite{simpson2015b} and the domain growth function \cite{le2018,ryabov2015,simpson2015c,yuste2016}. Here we are interested in the case where $D(C)$ is a non-negative function. For many applications, the population of interest exhibits a sharp front, which is not captured by the assumption of linear diffusion \cite{garduno1994,harris2004,mccue2019}. In this Letter, we derive and present exact solutions for diffusive processes on a growing domain for a class of degenerate (i.e. $D(0) = 0$) nonlinear diffusion models.
\section{Results}
\noindent We seek a series of transformations that yields a form of Equation \eqref{Eq:GoverningMain} that admits an exact solution. We begin by applying a boundary fixing transform \cite{johnston2023,simpson2015c}
\begin{equation}
\xi = \frac{r}{L(t)}L(0), \ -L(0) < \xi < L(0),
\label{Eq:SpaceTransform}
\end{equation}
which is the ratio of the domain size to the spatial variable. Applying this transformation yields
\begin{equation*}
\frac{\partial C(\xi,t)}{\partial t} =  \frac{1}{\xi^{d-1}}\bigg(\frac{L(0)}{L(t)}\bigg)^2 \frac{\partial}{\partial \xi}\bigg( \xi^{d-1}D(C) \frac{\partial C(\xi,t)}{\partial \xi}\bigg) - \frac{d}{L(t)}\frac{\text{d}L(t)}{\text{d}t}C(\xi,t).
\end{equation*} 
We see that we now have a dilution term for the mass when the domain is increasing in size. To remove the explicit time dependence in the diffusion term, we introduce a temporal scaling
\begin{equation*}
T(t) = \int_0^t \bigg(\frac{L(0)}{L(s)}\bigg)^2 \ \text{d}s,
\end{equation*}
which is the integral of the concentration-independent component of the (transformed) diffusivity, and gives
\begin{equation*}
\frac{\partial C(\xi,T)}{\partial T} = \frac{1}{\xi^{d-1}}\frac{\partial}{\partial \xi}\bigg(\xi^{d-1}D(C) \frac{\partial C(\xi,T)}{\partial \xi}\bigg) - d\frac{L(t)}{L(0)^2}\frac{\text{d}L(t)}{\text{d}t}C(\xi,T).
\end{equation*}
The time-dependent source term can be rescaled by defining
\begin{equation}
f(T) = d\frac{L(t)}{L(0)^2}\frac{\text{d}L(t)}{\text{d}t}, \qquad C(\xi,T) = U(\xi,T)\exp\bigg(-\int_0^T f(s) \ \text{d}s\bigg),
\label{Eq:DensityTransform}
\end{equation}
which gives
\begin{equation*}
\frac{\partial U(\xi,T)}{\partial T} = \frac{1}{\xi^{d-1}}\frac{\partial}{\partial \xi}\bigg( \xi^{d-1}D(C) \frac{\partial U(\xi,T)}{\partial \xi}\bigg).
\end{equation*}
The choice of the final transformation to rescale the $T$-dependency of the diffusion term depends on the form of the function $D(C)$. Here we choose $D(C) = mC^{m-1}$ and $m\neq1$, that is, the diffusivity in the porous medium equation when expressed in its divergence form\footnote{Noting that the porous medium equation is written as \begin{equation*} \frac{\partial C}{\partial t} = \Delta (C^m) = \nabla \cdot ( mC^{m-1}\nabla C ), \end{equation*}
so that we have $D(C) = mC^{m-1}$ and that the factor $m$ can be removed by rescaling time.}. As such, we have 
\begin{equation*}
\frac{\partial U(\xi,T)}{\partial T} = \frac{1}{\xi^{d-1}}\exp\bigg(-(m-1)\int_0^T f(s) \ \text{d}s\bigg)\frac{\partial}{\partial \xi}\bigg(m \xi^{d-1}U(\xi,T)^{m-1} \frac{\partial U(\xi,T)}{\partial \xi}\bigg).
\end{equation*}
Via a change of variable in the integral, we have
\begin{equation*}
 \exp\bigg(-(m-1)\int_0^T f(s) \ \text{d}s\bigg) = \Bigg(d\frac{L(0)}{L(t)}\Bigg)^{m-1},
\end{equation*}
and 
\begin{equation*}
\frac{\partial U(\xi,T)}{\partial T} = \frac{1}{\xi^{n-1}}\bigg(d\frac{L(0)}{L(t)}\bigg)^{m-1}\frac{\partial}{\partial \xi}\bigg(m\xi^{d-1}U(\xi,T)^{m-1} \frac{\partial U(\xi,T)}{\partial \xi}\bigg).
\end{equation*}
To obtain an exact solution, we seek a transformation such that the dependence on the ratio of domain sizes is not present in the scaled coordinates. We define $\tau(T)$ such that
\begin{equation*}
\frac{\text{d} \tau}{\text{d} T} = \frac{\text{d} \tau}{\text{d} t}\frac{\text{d} t}{\text{d} T} = \bigg(d\frac{L(0)}{L(t)}\bigg)^{m-1},
\end{equation*}
and hence
\begin{equation}
\tau(t) = \int_0^t  \bigg(n\frac{L(0)}{L(s)}\bigg)^{m+1} \ \text{d}s.
\label{Eq:TimeTransform}
\end{equation}
Applying this transformation, we ultimately obtain 
\begin{equation}
\frac{\partial U(\xi,\tau)}{\partial \tau} = \frac{1}{\xi^{d-1}}\frac{\partial}{\partial \xi}\bigg(m\xi^{d-1}U(\xi,\tau)^{m-1} \frac{\partial U(\xi,\tau)}{\partial \xi}\bigg),
\label{Eq:PME}
\end{equation}
which is the porous medium equation, for which there are several known exact and approximate solutions \cite{pattle1959,vazquez2007}. \\

For a Dirac delta initial condition, the Zeldovich-Kompaneets-Barenblatt solution to the porous medium equation on a $d$-dimensional infinite domain is
\begin{equation*}
U(\xi,\tau) = \tau^{-\alpha} \bigg(c - \frac{m-1}{2m}\gamma\xi^2\tau^{-2\gamma}\bigg)_+^{\frac{1}{m-1}} \qquad \text{where} \qquad \alpha = \frac{d}{d(m-1)+2}, \qquad \gamma = \frac{1}{d(m-1)+2},
\end{equation*}
and $c$ is an arbitrary constant related to the initial mass \cite{vazquez2007}. Solutions for the static (i.e. non-growing) domain case are shown in the Supplementary Information. Hence, from the transforms defined in Equations \eqref{Eq:SpaceTransform}, \eqref{Eq:DensityTransform} and \eqref{Eq:TimeTransform},
\begin{equation}
C(r,t) = d\frac{L(0)}{L(t)}\Bigg[\int_0^t  \bigg(d\frac{L(0)}{L(s)}\bigg)^{m+1} \ \text{d}s\Bigg]^{-\alpha} \Bigg(c - \frac{m-1}{2m}\gamma\bigg[r\frac{L(0)}{L(t)}\bigg]^2\bigg[\int_0^t  \bigg(d\frac{L(0)}{L(s)}\bigg)^{m+1} \ \text{d}s\bigg]^{-2\gamma}\Bigg)_+^{\frac{1}{m-1}},
\label{Eq:GeneralSolution}
\end{equation}
is the solution in the original $(r,t)$ coordinate system. We note that other exact or approximate solutions to the porous medium equation could also be transformed into solutions in the growing domain coordinate system. Further, we could choose a different diffusivity function and, provided that a $\tau(T)$ transformation exists such that we obtain an equation that admits an exact solution, we can determine an exact solution in the original coordinate system. \\

Equation \eqref{Eq:GeneralSolution} is defined for an infinite domain. However, the standard approach for diffusive processes on growing domains is to consider a finite domain where any mass that reaches the boundary is immediately removed from the system. We therefore impose the boundary condition $C(-L(t),t) = C(L(t),t) = 0$. \\

If we consider a finite static domain, the population will eventually reach the boundary and the solution will no longer satisfy the boundary conditions. However, the porous medium equation admits solutions with compact support \cite{vazquez2007}. Therefore, for growing domains, we are interested in addressing the question of whether the population spreads faster than the growth rate of the domain. We define $r^*(t)$ as the width of the support for the solution at time $t$, that is, that $C(r,t) = 0$ for $r \geq r^*(t)$. If $L(t) > r^*(t) \ \forall \, t$ then the population will never reach the domain boundary, the boundary conditions will be satisfied, and hence the solution is valid. Further, provided the population does not reach the domain boundary, the solution also satisfies homogeneous Neumann and periodic boundary conditions. Equation \eqref{Eq:PME} is invariant in $\tau$ (but not $t$). Accordingly, we are not restricted to a Dirac delta initial condition; instead, we could consider a shifting in $\tau$ such that the initial condition is any solution $U(\xi,\tau)$. However, here we focus on the case where $C(r,0) = F(r) = \delta(r)$. \\

Two common choices of domain growth functions are linear growth, where $L(t) = L(0) + \beta t$, and exponential growth, where $L(t) = L(0)\exp(\beta t)$ \cite{johnston2023,simpson2015c}. In both cases, $\beta > 0$. The corresponding time transformations are
\begin{equation*}
\tau(t) = \int_0^t  \bigg(d\frac{L(0)}{L(0)+\beta s}\bigg)^{m+1} \ \text{d}s = \frac{d^{m+1}L(0)\Big(1 - \big[\frac{L(0)}{L(0)+\beta t}\big]^m\Big)}{\beta m},
\end{equation*}
and
\begin{equation*}
\tau(t) = \int_0^t  \Big(d\exp(-\beta s)\Big)^{m+1} \ \text{d}s = \frac{d^{m+1}\big(1 - \exp[-\beta(m+1)t]\big)}{\beta(m+1)}.
\end{equation*}
For $m > 1$, the mass that evolves according to the porous medium equation exhibits compact support \cite{vazquez2007}. The width of this support, in the original coordinates, is
\begin{equation}
r^*(t) = \bigg(\frac{2cm}{\gamma(m-1)}\tau(t)^{2\gamma}\bigg)^{1/2}\frac{L(t)}{L(0)},
\label{Eq:Support}
\end{equation}
for the relevant choice of domain growth function, $L(t)$. We see that the width of the support scales with the ratio of the current domain size to the original domain size. We require that $r^*(t) \leq L(t) \ \forall t$ for the boundary conditions to be satisfied by the solution. For a linearly growing domain this is satisfied for all $t$ if
\begin{equation*}
\frac{\beta m}{d^{m+1}L(0)}\bigg(\frac{\gamma(m-1)L(0)^2}{2cm}\bigg)^{\frac{1}{2\gamma}} \geq 1,
\end{equation*}
and, otherwise, will be satisfied for a finite interval of $0 \leq t \leq t^*$ where
\begin{equation}
t^* = \frac{L(0)}{\beta}\Bigg(\Bigg[1 - \frac{\beta m}{d^{m+1}L(0)}\bigg(\frac{\gamma(m-1)L(0)^2}{2cm}\bigg)^{\frac{1}{2\gamma}}\Bigg]^{\frac{-1}{m}}-1\Bigg).
\label{Eq:CritTimeLinear}
\end{equation}

For an exponentially growing domain, the boundary conditions are satisfied for all $t$ if
\begin{equation*}
\frac{\beta (m+1)}{d^{m+1}}\bigg(\frac{\gamma(m-1)L(0)^2}{2cm}\bigg)^{\frac{1}{2\gamma}} \geq 1,
\end{equation*}
and, otherwise, will be satisfied for a finite interval of $0 \leq t \leq t^*$ where
\begin{equation}
t^* = \frac{1}{\beta} \ln\Bigg(\Bigg[1 - \frac{\beta (m+1)}{d^{m+1}}\bigg(\frac{\gamma(m-1)L(0)^2}{2cm}\bigg)^{\frac{1}{2\gamma}}\Bigg]^{\frac{-1}{m+1}}\Bigg).
\label{Eq:CritTimeExp}
\end{equation}
An increase in either the domain growth rate or the initial size of the domain results in an increase in $t^*$. Moreover, these conditions allow us to identify a critical growth rate $\beta^*$ for a specific initial domain length (or, equally, a critical initial domain length for a specific growth rate) for a specified diffusivity function and dimension, above which none of the population will reach the moving boundary and below which at least a fraction of the population will reach the moving boundary. For a linearly growing domain,
\begin{equation}
\beta^* = \frac{d^{m+1}L(0)}{m}\bigg(\frac{\gamma(m-1)L(0)^2}{2cm}\bigg)^{\frac{-1}{2\gamma}}, 
\label{Eq:CritGrowthRateLinear}
\end{equation}
and, for an exponentially growing domain,
\begin{equation}
\beta^* = \frac{d^{m+1}}{m+1}\bigg(\frac{\gamma(m-1)L(0)^2}{2cm}\bigg)^{\frac{-1}{2\gamma}}.
\label{Eq:CritGrowthRateExp}
\end{equation}

\begin{figure}[t!]
\begin{center}
\includegraphics[width=1.0\textwidth]{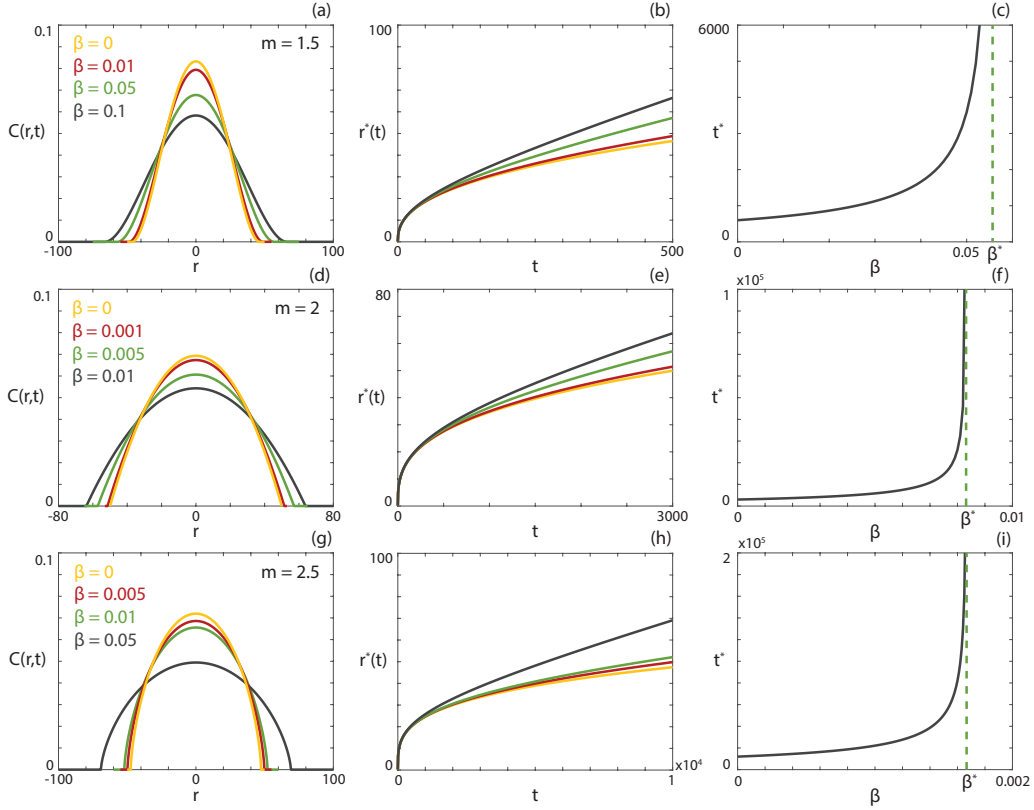}
\end{center}
\caption{(a),d),(g) Solution profiles (Equation \eqref{Eq:GeneralSolution}), (b),(e),(h) support widths (Equation \eqref{Eq:Support}) and (c),(f),(i) critical times (Equation \eqref{Eq:CritTimeLinear}) for different linear domain growth rates with (a)-(c) $m = 1.5$, (d)-(f) $m = 2$ and (g)-(i) $m = 2.5$ for a domain with $L(0) = 50$. Solution profiles are presented at (a) $t = 500$, (d) $t = 3000$ and (h) $t = 1\times 10^4$. The critical domain growth rates are (a)-(c) $\beta^* \approx 5.57 \times 10^{-2}$, (d)-(f) $\beta^* \approx 8.33 \times 10^{-3}$, (g)-(i) $\beta^* \approx 1.67 \times 10^{-3}$. Results are presented for $c = 1$ and $d = 1$.}
\label{F1}
\end{figure}

We present solution profiles, $C(r,t)$, the width of the support, $r^*(t)$, and the critical time, $t^*$, for a range of $\beta$ and $m$ values for the case of linear domain growth in Fig. \ref{F1}. Similar results for exponential domain growth are presented in the Supplementary Information (SI Fig. 3). The solution profiles are presented at $t < t^*$ in the cases where $\beta < \beta^*$, so the profiles obey the boundary conditions in each case. We observe that as the domain growth increases, the population spreads more rapidly, as expected. As we increase $m$, we see that the solution profiles change from exhibiting negative curvature as $C(r,t) \to 0^+$ (Fig. \ref{F1}(a)) to exhibiting positive curvature as $C(r,t) \to 0^+$ (Fig. \ref{F1}(g)). This feature of the solution profiles is consistent with solutions arising from the porous medium equation on a static domain \cite{vazquez2007}. Intuitively, in Figs. \ref{F1}(c),(f),(i), we observe that $t^* \to \infty$ as $\beta \to \beta^*$. \\

\begin{figure}
\begin{center}
\includegraphics[width=0.8\textwidth]{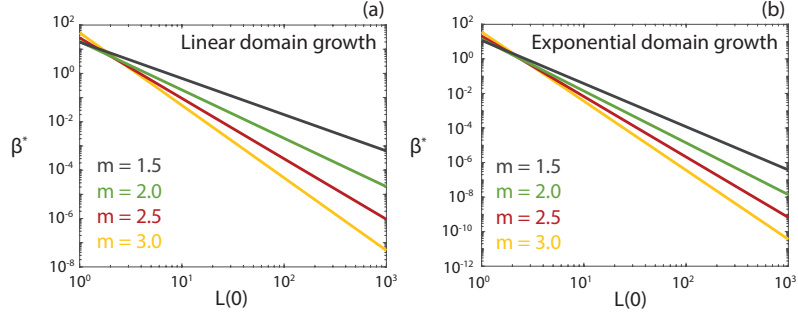}
\end{center}
\caption{Critical domain growth rates for (a) linear domain growth (Equation \eqref{Eq:CritGrowthRateLinear}) and (b) exponential domain growth (Equation \eqref{Eq:CritGrowthRateExp} for $m = 1.5$ (grey), $m = 2$ (green), $m = 2.5$ (red) and $m = 3$ (yellow) and a range of initial domain sizes. Results are presented for $c = 1$ and $d = 1$.}
\label{F2}
\end{figure}

We highlight how the critical domain growth rate changes with the initial domain size in Figure \ref{F2} for both linearly and exponentially growing domains and a suite of $m$ values. In both cases, we observe that for small initial domain sizes, higher $m$ values require more rapid domain growth for the solution to be valid for all time. However, as $L(0)$ increases, lower $m$ values require higher domain growth rates than higher $m$ values. This is because for the Dirac delta initial condition, the population support increases more rapidly at early time for higher $m$ values, as the population is highly concentrated. As the population disperses, the evolution of the support becomes faster for lower $m$ values (Figs. \ref{F1}(b),(e),(h) and Equation \eqref{Eq:Support}). 

\section{Discussion and Conclusions}
\noindent In this Letter we have presented new exact solutions to a model of a nonlinear diffusive process on growing domains. Growing domains occur throughout nature \cite{bers2002,landman2007,riess1998,shraiman2005}, and there is significant interest in understanding how the combination of diffusion and domain growth manifests in the spread the population \cite{landman2007}. Previous investigations have primarily focused on linear diffusive processes \cite{johnston2023,le2018,ryabov2015,simpson2015b,simpson2015c,yuste2016}. However, in many applications, the population of interest exhibits a sharp front, which is not captured by linear diffusion \cite{garduno1994,mccue2019}. Here we extend the transformation approach considered in prior studies \cite{johnston2023,simpson2015b} such that the model of nonlinear diffusion on a growing domain can be transformed into a nonlinear diffusion equation on a static domain, which admits exact solutions for certain choices of diffusivity functions and boundary conditions \cite{vazquez2007}. Upon inverting the transformations, we obtain exact solutions that describe the spread of a population undergoing nonlinear diffusion on a growing domain. The exact solutions allow us to probe and analyse the relationship between model parameters and the spreading rate, and identify critical domain growth rates and timescales for the solutions to be valid. \\

We have restricted our focus to a diffusive population. However, it would be straightforward to include a linear growth term and use a similar set of transformations to obtain an exact solution. While the assumption of linear growth is overly simplistic for many populations where crowding effects are important \cite{johnston2017}, this can be considered as a linearisation of a higher order growth term that is valid at low population densities. In previous studies, one benefit of the exact solutions is that relevant statistics such as the splitting and survival probabilities can be calculated exactly \cite{johnston2023,simpson2015b,simpson2015c,yuste2016}. Here, for the solution to be valid, we require that the population does not reach the domain boundaries and hence we are only able to describe the case where the survival probability is exactly one. Other investigations have considered stochastic models, such as a lattice-based random walk on a growing lattice, and the relationship between such models and continuum models of linear diffusive processes on growing domains \cite{baker2010,yates2012,yates2014}. While the impact of the choice of the specific stochastic domain growth model has been explored in depth \cite{yates2012}, the details of a lattice-based random walk that would faithfully describe the behaviour of a population with a sharp front remains unclear. Accordingly, it would be of interest to determine the random walk process that is (approximately) consistent with the model examined in this Letter.  \\

\noindent \textit{Data Accessibility.} The code used to generate the results presented here can be found on Github at: \\
https://github.com/DrStuartJohnston/nonlinear-diffusion-growing-domains. \\

\noindent \textit{Authors' Contributions.} \textbf{STJ}: Conceptualisation, Formal Analysis, Investigation, Methodology, Resources, Software, Validation, Project Administration, Visualisation, Writing - Original Draft, Writing - Review \& Editing. \textbf{MJS}: Conceptualisation, Formal Analysis, Investigation, Methodology, Validation, Writing - Review \& Editing. \\

\noindent \textit{Competing Interests.} The authors declare that they have no competing interests. \\

\noindent \textit{Funding.} This work was in part supported by the Australian Research Council (STJ: DP230100380, MJS: DP200100177).

\bibliographystyle{elsarticle-num}

\section*{Supplementary Information}
\begin{figure}[h!]
\begin{center}
\includegraphics[width=0.8\textwidth]{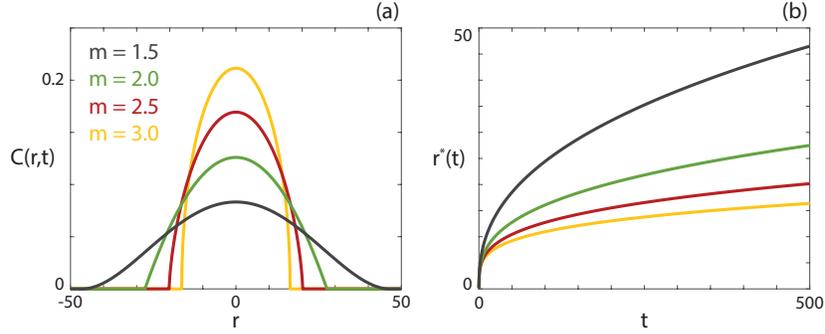}
\end{center}
\caption{Impact of the polynomial order of the diffusivity function on the solution for a static domain. (a) Solution profiles (Equation (7)) and (b) support widths (Equation (9)) for a static domain with $m = 1.5$ (grey), $m = 2$ (green), $m = 2.5$ (red) and $m = 3$ (yellow). Solution profiles are presented at $t = 500$. Results are presented for $c = 1$ and $d = 1$.}
\label{SI_F1}
\end{figure}

\begin{figure}[h!]
\begin{center}
\includegraphics[width=0.8\textwidth]{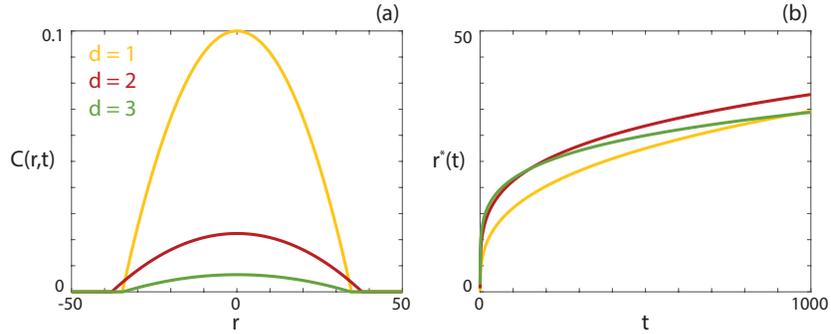}
\end{center}
\caption{Impact of the domain dimension on the solution for a static domain. (a) Solution profiles (Equation (7)) and (b) support widths (Equation (9)) for a static domain with $d = 1$ (yellow), $d = 2$ (red) and $d = 3$ (green). Solution profiles are presented at $t = 500$. Results are presented for $c = 1$ and $m = 2$.}
\label{SI_F2}
\end{figure}

\begin{figure}[h!]
\begin{center}
\includegraphics[width=1.0\textwidth]{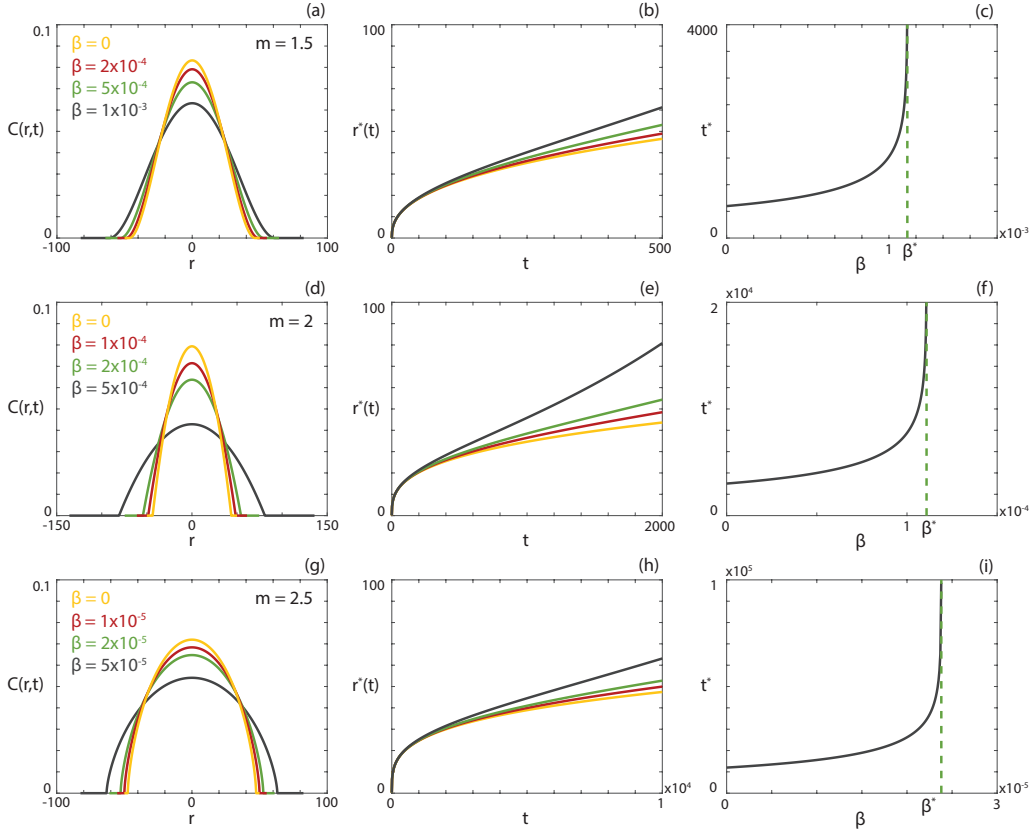}
\end{center}
\caption{(a),d),(g) Solution profiles (Equation (7)), (b),(e),(h) support width (Equation (10)) and (c),(f),(i) critical times (Equation (11)) for different combinations of exponential domain growth rates for (a)-(c) $m = 1.5$, (d)-(f) $m = 2$ and (g)-(i) $m = 2.5$ for a domain with $L(0) = 50$. Solution profiles are presented at (a) $t = 500$, (d) $t = 2000$ and (h) $t = 1\times 10^4$. The critical domain growth rates are (a)-(c) $\beta^* \approx 6.68 \times 10^{-4}$, (d)-(f) $\beta^* \approx 1.11 \times 10^{-4}$, (g)-(i) $\beta^* \approx 2.38 \times 10^{-5}$.}
\label{SI_F3}
\end{figure}

Here we present additional results to the main manuscript. In Figure \ref{SI_F1}, we highlight the impact that the polynomial order of the diffusivity function (i.e. $m$) has on the solution profile for a static domain. As in the main manuscript, we see the classic behaviour of an increasingly steep front for increasing $m$. In Figure \ref{SI_F2}, we show how changing the dimensionality of the model impacts the solution profile and the rate of spread of the population. In Figure \ref{SI_F3}, we present solution profiles, the width of the support, and the critical time, for a range of $\beta$ and $m$ values for the case of exponential domain growth.

\end{document}